\documentclass[fleqn,usenatbib]{mnras}
\usepackage{newtxtext,newtxmath,hyperref}

\DeclareRobustCommand{\VAN}[3]{#2}
\let\VANthebibliography\thebibliography
\def\thebibliography{\DeclareRobustCommand{\VAN}[3]{##3}\VANthebibliography}

\usepackage[T1]{fontenc}
\usepackage{ae,aecompl}
\usepackage[dvipdfmx]{graphicx}	
\usepackage{amsmath}	
\usepackage{amssymb}	
\usepackage{multirow,cases,colortbl}
\usepackage{url}
\usepackage{bm}
\usepackage{lscape}
\usepackage{ulem,color}

\newcommand{\Msun}{\ensuremath{\mathrm{M}_\odot}}
\newcommand{\Rd}{\ensuremath{R_{\rm{d}}}}
\newcommand{\Rc}{\ensuremath{R_{\rm{c}}}}

\newcommand{\vd}{\ensuremath{v_{\rm{d}}}}
\newcommand{\vc}{\ensuremath{v_{\rm{c}}}}

\newcommand{\Kes}{\ensuremath{\kappa_{\rm{es}}}}
\newcommand{\Ka}{\ensuremath{\kappa_{\rm{a}}}}
\newcommand{\Ko}{\ensuremath{\kappa_{\rm{0}}}}
\newcommand{\rhod}{\ensuremath{\rho_{\rm{d}}}}
\newcommand{\rhoc}{\ensuremath{\rho_{\rm{c}}}}

\newcommand{\Mej}{\ensuremath{{M}_{\rm{ej}}}}
\newcommand{\Ekin}{\ensuremath{{E}_{\rm{kin}}}}
\newcommand{\SB}{\ensuremath{{\sigma}_{\rm{SB}}}}

\title[Ejected mass of optical TDEs]{Limits on mass outflow from optical tidal disruption events}

\author[Matsumoto \& Piran]{Tatsuya Matsumoto$^{1,2,3}$\thanks{E-mail: tatsuya.matsumoto@mail.huji.ac.il}\thanks{JSPS Research Fellow} and Tsvi Piran$^{1}$
\\
$^{1}$Racah Institute of Physics, Hebrew University, Jerusalem, 91904, Israel\\
$^{2}$Research Center for the Early Universe, Graduate School of Science, University of Tokyo, Tokyo 113-0033, Japan\\
$^{3}$Department of Physics, Graduate School of Science, University of Tokyo, Tokyo 113-0033, Japan\\
}

\pubyear{2020}

\begin{document}
\label{firstpage}
\pagerange{\pageref{firstpage}--\pageref{lastpage}}
\maketitle

\begin{abstract}
The discovery of optical/UV tidal disruption events (TDEs) was  surprising. The expectation was that, upon returning to the pericenter, the stellar-debris stream will form a compact disk that will emit soft X-rays. Indeed the first TDEs were discovered in this energy band. A common explanation for the optical/UV events is that surrounding optically-thick matter reprocesses the disk's X-ray emission and emits it from a large photosphere. If accretion follows the super-Eddington mass infall rate it would inevitably result in an energetic outflow, providing naturally the reprocessing matter. We describe here a new method to estimate, using the observed luminosity and temperature, the mass and energy of outflows from optical transients. When applying this method to a sample of supernovae our estimates are consistent with a more detailed hydrodynamic modeling. For the current sample of a few dozen optical TDEs the observed luminosity and temperature imply outflows that are significantly more massive than typical stellar masses, posing a problem to this common reprocessing picture.
\end{abstract}

\begin{keywords}
radiation mechanisms: thermal, (stars:) supernovae:general, transients: supernovae, transients: tidal disruption events
\end{keywords} 

\section{Introduction}
A tidal disruption event (TDE) occurs when a star approaches a supermassive black hole (BH) in a galactic center and reaches the tidal radius where the tidal force is strong enough to overcome the star's self-gravity. 
After the disruption, about half of the stellar debris is unbound.  The other bound half falls back to the BH with a fallback rate $\dot{M}_{\rm fb}\propto t^{-5/3}$ \citep{Rees1988,Phinney1989}.
Upon returning to the pericenter, if the fallback stream circularized rapidly it will form a compact accretion disk. A disk of this size will produce soft X-ray/UV emissions.
About ten TDEs, whose location is consistent with the center of their host galaxies and  their light curve decays as the accretion rate $\propto t^{-5/3}$, have been detected in the soft X-ray band by the ROSAT survey \citep{Brandt+1995,Grupe+1995,Bade+1996,Grupe+1999,Komossa&Bade1999,Komossa&Greiner1999,Greiner+2000}, XMM-Newton \citep{Esquej+2007,Esquej+2008}, and Chandra (\citealt{Maksym+2013,Donato+2014}, see \citealt{Komossa2015} for a review of these early observations).

A different class of TDEs, hereafter denoted ``optical TDEs'',  was discovered in the last decade in optical and UV bands. Like the earlier X-ray TDEs these events are also located in galactic nuclear regions and their light curves decline roughly as $t^{-5/3}$.
A few dozen such events have been detected by wild-field high-cadence optical/UV surveys, such as GALEX \citep{Gezari+2006,Gezari+2008}, SDSS \citep{vanVelzen+2011}, pan-STARRS \citep{Gezari+2012,Chornock+2014}, PTF \citep{Arcavi+2014}, iPTF \citep{Blagorodnova+2017,Blagorodnova+2019,Hung+2017}, and ASASSN \citep{Holoien+2014,Holoien+2016,Holoien+2016b}.
Recently, ZTF almost doubled the sample size \citep{vanVelzen+2020,Nicholl+2020}.

Common characteristics of optical TDEs that are most relevant to this work are \citep{Hung+2017,vanVelzen+2020,vanVelzen+2020b}: (i)  a peak bolometric luminosity  $ L\sim10^{44}\,\rm erg\,s^{-1}$, (ii)  a typical duration longer than $t \gtrsim100\,\rm days$, (iii) a black-body temperature of $T\simeq3\times10^{4}\,\rm K$ that does not vary much during the observation. 
(iv) spectra dominated by a blue continuum component. Some events show broad emission lines corresponding to the velocity of $v\lesssim10^4\,\rm km\,s^{-1}$ \citep{Arcavi+2014}.
(v) a  total emitted energy $L t \sim10^{51}\, \rm erg$ that is much lower than the expected total energy from efficient accretion of a solar mass onto a BH ($\sim 10^{53}$ erg) or even from the energy required to circularize the flow onto an accretion disk ($\sim 10^{52}$ erg). This is the so-called ``inverse energy crisis''  \citep{Piran+2015,Stone&Metzger2016,Lu&Kumar2018}.

These properties of optical TDEs are inconsistent with the classical picture in which soft X-rays from an accretion disk are expected to dominate the emission.
Instead, it was proposed that the emission is reprocessed by surrounding matter.
\cite{Strubbe&Quataert2009,Lodato&Rossi2011} proposed an outflow launched from small radii (probably a super-Eddington disk wind) radiating away its thermal energy. 
Later on others \citep{Metzger&Stone2016,Roth+2016,Roth&Kasen2018,Dai+2018,Lu&Bonnerot2020,Piro&Lu2020} proposed that an outflow (either a super-Eddington disk wind or the result of shocks in stream-stream or stream-disk collisions) expands, remains optically thick, and reprocesses the ionizing continuum from the inner accretion disk. 
We denote these as ``reprocessing-outflow" models.
Since the fallback rate is much larger than the Eddington rate, super-Eddington emission \citep{Loeb&Ulmer1997,Krolik&Piran2012} and an outflow from the disk \citep[e.g.,][]{Blandford&Begelman1999} are naturally expected.
The expanding material surrounds the system and reprocesses the soft X-rays emitting photons in optical/UV band if thermalization is efficient \citep{Roth+2016,Roth&Kasen2018}.
\cite{Metzger&Stone2016} have suggested that the outflow carries out a significant fraction of the infalling mass, reducing the mass accreted by the BH and hence the energy generation rate. 
Alternatively, the outflow can carry out the excess energy in the form of kinetic energy. This would have resolved the ``inverse energy crisis'' \citep{Piran+2015,Stone&Metzger2016,Lu&Kumar2018}.

An alternative model \citep{Piran+2015,Krolik+2016} suggests that  the observed optical emission is generated by interactions between the bound stellar debris taking place around the apocenter.
This model follows the simulations of \cite{Shiokawa+2015} who have shown that the fallback stream passes the pericenter without forming a disk and it collides with the debris near the apocenter.
Heated by shocks, the interacting part powers the observed optical emission \citep[see also][]{Svirski+2017,Ryu+2020}. In this case the accretion onto the BH is delayed and is possibly inefficient. 

In this work we focus on the former scenario in which we expect an outflow that is relevant to the reprocessing process. 
Following recent works by \cite{Shen+2015} and \cite{Piro&Lu2020}, we analyze the condition within the emitting region and estimate the ejecta mass involved in the optical TDEs. We then impose the condition that this mass cannot exceed the disrupted stellar mass, which is of order of a solar mass and explore its implications on the reprocessing model.
We organize the paper as follows.
In \S \ref{sec method}, we develop the method to estimate the ejecta mass of optical transients in a general quasi-spherical optically thick situation using the observed luminosity and temperature.
In \S \ref{sec sn}, we apply this method to supernovae (SNe) and confirm that we can estimate the ejecta mass with a good accuracy.
In \S \ref{sec tde}, we calculate the ejecta mass of available optical TDEs by assuming that the emission is powered by spherically expanding wind. We summarize our result and their implications in \S \ref{sec summary}.

\section{Method}\label{sec method}
We construct a framework to estimate ejecta mass of optical transients assuming that (i) they expand quasi-spherically and (ii) they are optically thick.  The observed photons are thermal and diffuse out of the ejecta.
In the context of optical TDEs this situation arises within the ``reprocessing-outflow'' model.
We note that our framework is  relevant not only to explosive phenomena like SNe and other transients \citep{Piro&Lu2020,Uno&Maeda2020}, but also to quasi-steady-state configurations as was discussed in \cite{Shen+2015} for the ultraluminous X-ray source M101 X-1.

We begin defining two critical radii that determine the observables \citep[e.g.,][]{Nakar&Sari2010,Shen+2015}. The first  is the diffusion radius, $\Rd$, (denoted photon trapping radius, $R_{\rm trap}$, by \citealt{Shen+2015}) above which the photon diffusion time is shorter than the dynamical time and photons can freely escape from the ejecta. Using the total optical depth we write this condition as
\begin{align}
\tau(\Rd)\equiv\int_{\Rd}^\infty (\Kes+\Ka)\rho dR=\frac{c}{\vd},
    \label{eq tau}
\end{align}
where $\kappa_{\rm es}$ and $\kappa_{\rm a}$ are the Thomson (electron scattering) and absorption opacities, $\rho$ is the density, $c$ is the speed of light, and $\vd$ is the ejecta velocity at $\Rd$.

The color radius, $\Rc$, (denoted thermalization radius, $R_{\rm th}$, by \citealt{Shen+2015}) is the location where the photons' last absorption occurs (namely, photons are in thermal equilibrium with the gas within $\Rc$). 
This radius is defined by the effective optical depth as \citep{Rybicki&Lightman1979}:
\begin{align}
\tau_{\rm eff}(\Rc)\equiv\int_{\Rc}^\infty \sqrt{\kappa_{\rm a}(\kappa_{\rm es}+\kappa_{\rm a})}\rho dR = 1 \ .
    \label{eq tau_eff}
\end{align}
Strictly speaking, photons with different energies will have different effective optical depths. We define the color radius as the radius corresponding to the observed color temperature \citep[see][for details]{Shen+2015,Lu&Bonnerot2020}. In order to derive this radius, we use for $\Ka$ the Planck-mean absorption opacity.

\begin{figure}
\begin{center}
\includegraphics[width=85mm, angle=0]{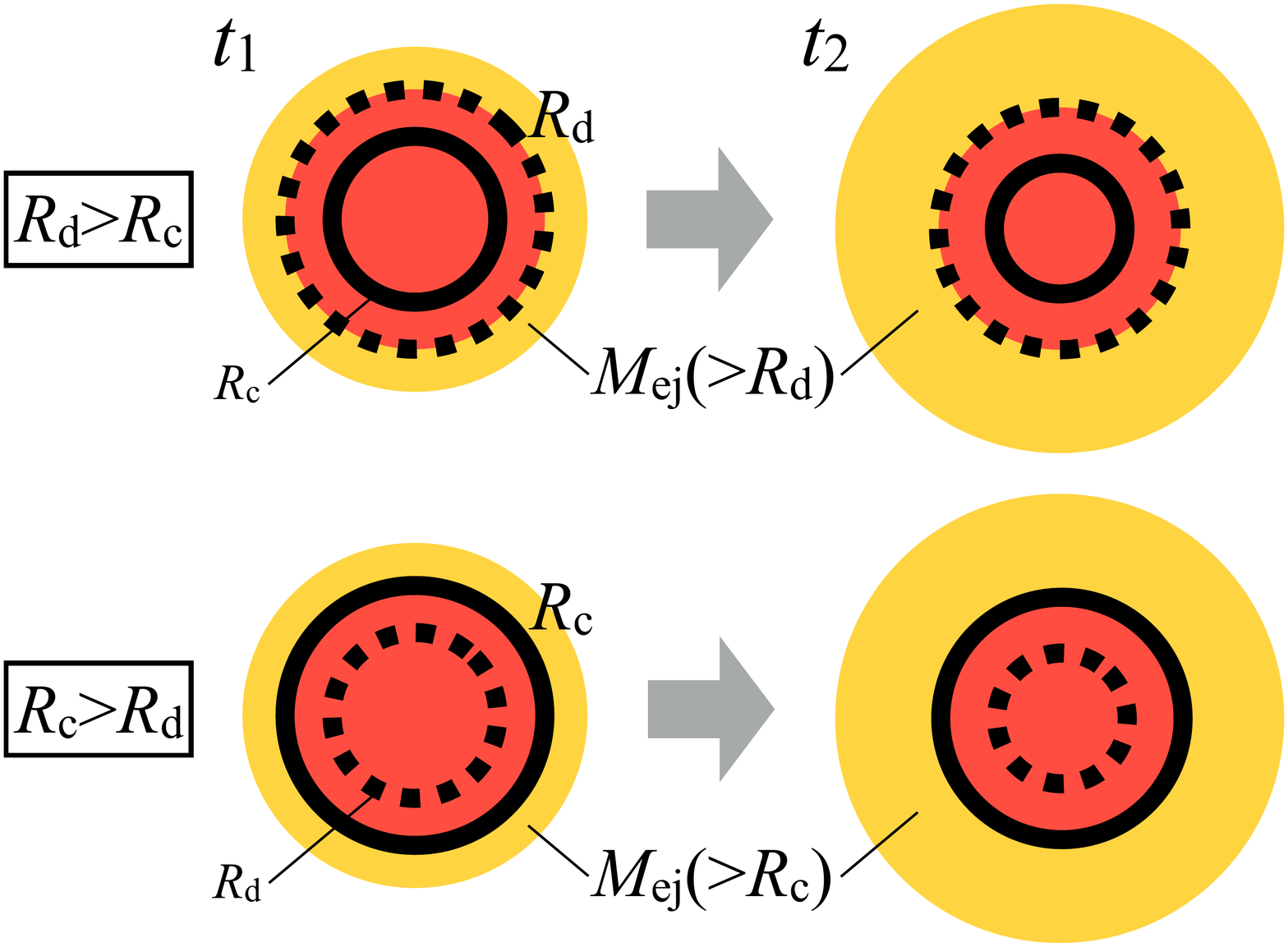}
\caption{A schematic picture of system at two time $t_1< t_2$ for the two different cases. ({\bf Top})  $\Rd>\Rc$, photons are trapped within the ejecta up to the diffusion radius ($\Rd$, dotted curve) and escape from the regions beyond $\Rd$ (marked by yellow color). ({\bf Bottom}) $\Rc>\Rd$, photons diffuse out of the ejecta beyond $\Rd$, but they are still in thermal equilibrium with the gas up to the color radius ($\Rc$, solid curve). Outside of $\Rc$ (also  marked by yellow color), the photons decouple from the gas.
As time progresses, both radii $R_{\rm d}$ and $\Rc$ shrink in the Lagrangian (mass) coordinate.
Integrating over time the mass outflow rate $\dot{M}$ at $\Rd$ and $\Rc$, we estimate the ejecta mass that crossed $\Rd$, $\Mej(>\Rd)$ for the first case and the mass that crossed $\Rc$, $\Mej(>\Rc)$ for the second.
}
\label{fig picture}
\end{center}
\end{figure}

The system has two different physical situations depending on whether $\Rd>\Rc$ or $\Rc>\Rd$.
In the former case, photons are trapped within the ejecta and advected up to $\Rd$. The observed luminosity is given by the diffusion luminosity there. 
The observed color temperature is given by the photon temperature at $\Rd$,
which is the same as the gas temperature.\footnote{Note that beyond $\Rc$ the photons and the gas cool adiabatically as $T\propto\rho^{1/3}$ and $T_{\rm g}\propto\rho^{2/3}$, respectively. However, Compton heating will heat the gas to the same temperature as the radiation in radiation-pressure-dominated ejecta.} When $\Rc>\Rd$, photons diffuse out from $\Rd$ but they are still thermally coupled to the gas. The observed color temperature is determined by the radiation temperature at $\Rc$.
Fig. \ref{fig picture} depicts a schematic picture of the system we consider for the cases of $\Rd>\Rc$ and  $\Rd<\Rc$.

Using the observed luminosity $L$, temperature $T$, and the outflow velocity (when it is available) we can estimate the conditions at $R={\rm max}(\Rd,\Rc)$ and using them we calculate the mass outflow rate passing through this radius:
\begin{align}
\dot{M}=4\pi R^2\rho \biggl(v-\frac{dR}{dt}\biggl) \ ,
   \label{eq mdot}
\end{align}
where $v$ is the ejecta velocity at $R$.
The last term $dR/dt$ arises because the radii $\Rd$ and $\Rc$ move in both the Eulerian and Lagrangian (mass) coordinates.
Usually, as the ejecta expand, these radii recede in the mass coordinate.
Integrating the mass outflow rate over time we calculate the total ejecta mass above this radius.

To determine $\Rd$ and $\Rc$ we use the simplified forms of Eqs. \eqref{eq tau} and \eqref{eq tau_eff}. For the absorption opacity, we adopt the Kramers' opacity $\Ka=\Ko\rho T_{\rm g}^{-7/2}$ for TDEs, where $\Ko$ is a constant that depends on the composition and  $T_{\rm g}$  is the gas temperature. Since the absorption opacity is always smaller than the Thomson opacity ($\kappa_{\rm a}\ll\Kes$), and as long as the density profile is steeper than $\rho\propto r^{-1}$,  we can approximate the optical depths in Eqs. \eqref{eq tau} and \eqref{eq tau_eff} and obtain: 
\begin{align}
&\tau(\Rd)\simeq\Kes\rhod \Rd=c/\vd \ ,
    \label{eq def r_d}\\
&\tau_{\rm eff}(\Rc)\simeq\sqrt{\kappa_{\rm es}\kappa_{\rm a}}\rhoc \Rc=1 \ ,
    \label{eq def r_c}
\end{align}
where $\rhod$ and $\rhoc$ are the density at $\Rd$ and $\Rc$, respectively.
For TDEs in which the chemical abundance is solar and temperatures are $\simeq3\times10^4\,\rm K$ we use $\Kes=0.35\,\rm cm^2\,g^{-1}$ and $\Ko=4\times10^{25}$. We also use this values as fiducial values in the rest of this section.
When we analyze type Ic SNe in \S \ref{sec sn}, whose temperatures are much lower and whose ejecta are  hydrogen and helium free, we use $\Kes=0.07\,\rm cm^2\,g^{-1}$ and a constant absorption opacity of $\Ka=0.01\,\rm cm^2\,g^{-1}$ (see Appendix \ref{appendix a} for details of our opacity prescription).

The  mass outflow rate (Eq. \ref{eq mdot}) depends on whether $\Rd > \Rc$ or $\Rc > \Rd$.
In the former case, the observed luminosity and color temperature are determined at $\Rd$.
The bolometric luminosity is given by the diffusion approximation at $\Rd$:
\begin{align}
L=-\frac{4\pi \Rd^2 ac}{3\Kes\rhod}\frac{dT^4}{dR}\simeq\frac{4\pi \Rd^2 a\vd T^4}{3} \ ,
\end{align}
where $a$ is the radiation constant and we used Eq. \eqref{eq def r_d}.
Solving this equation for $\Rd$, we obtain
\begin{align}
\Rd&\simeq\biggl(\frac{3c}{2^4\pi\SB}\biggl)^{1/2}L^{1/2}T^{-2}\vd^{-1/2} \ ,
	\label{eq r_d}
\end{align}
where $\sigma_{\rm SB}=ac/4$ is the Stefan-Boltzmann constant.
Combining this equation with Eq. \eqref{eq def r_d}, we  estimate $\rho_{\rm d}$:
\begin{align}
\rhod&\simeq\biggl(\frac{2^4\pi c\SB}{3\Kes^2}\biggl)^{1/2}L^{-1/2}T^{2}\vd^{-1/2} \ .
	\label{eq rho_d}
\end{align}
Combined together, the mass outflow rate at $\Rd$ is given by
\begin{align}
\dot{M}_{\rm d}&\simeq\biggl(\frac{3\pi c^3}{\Kes^2\SB}\biggl)^{1/2}L^{1/2}T^{-2}\vd^{-1/2}f_{\rm d}
	\label{eq mdot d}\\
&\simeq8.3\times10^{-1}{\,\Msun\,\rm day^{-1}\,}L_{44}^{1/2}T_{4}^{-2}v_{\rm d,9}^{-1/2}f_{\rm d} \ ,\nonumber\\
f_{\rm d}&=1-\frac{\frac{d\Rd}{dt}}{\vd}=1-\frac{\Rd}{\vd}\biggl(\frac{1}{2}\frac{d\ln L}{dt}-2\frac{d\ln T}{dt}-\frac{1}{2}\frac{d\ln \vd}{dt}\biggl) \ ,
\end{align}
where we use the convention of $Q_x = Q/10^x$ (in cgs units).

We turn to the case of $\Rc>\Rd$, in which  photons are coupled to the gas up to $\Rc$ and the observed photon temperature equals to the  gas temperatures at $\Rc$, $T=T_{\rm g}$.
We can still use the diffusion approximation at $\Rc$\footnote{When $\tau_{\rm eff}(\Rc)=1$, we have $\tau(\Rc)=(\Kes/\kappa_{\rm a})^{1/2}>1$.} and the bolometric luminosity is given by
\begin{align}
L=-\frac{4\pi \Rc^2 ac}{3\Kes\rhoc}\frac{dT^4}{dR}\sim\frac{4\pi \Rc ac T^4}{3\Kes\rhoc} \ .
\end{align}
Combining this equation with Eq. \eqref{eq def r_c}, we estimate $\Rc$ and $\rhoc$:
\begin{align}
\rhoc&=\biggl(\frac{2^8\pi^2\SB^2}{3^2\Kes^3\Ko}\biggl)^{1/5}L^{-2/5}T^{23/10} \ ,
	\label{eq rho_c}\\
\Rc&=\biggl(\frac{3^3\Kes^2}{2^{12}\pi^3\Ko\SB^3}\biggl)^{1/5}L^{3/5}T^{-17/10} \ .
	\label{eq r_c}
\end{align}
Finally, the mass outflow rate at $\Rc$ is given by
\begin{align}
\dot{M}_{\rm c}&\simeq\biggl(\frac{3^4\pi\Kes}{2^6\Ko^3\SB^4}\biggl)^{1/5}L^{4/5}T^{-11/10}\vc f_{\rm c}
	\label{eq mdot c}\\
&\simeq3.2\times10^{-1}{\,\Msun\,\rm day^{-1}\,}L_{44}^{4/5}T_{4}^{-11/10}v_{\rm c,9}f_{\rm c} \ ,\nonumber\\
f_{\rm c}&=1-\frac{\frac{d\Rc}{dt}}{\vc}=1-\frac{\Rc}{\vc}\biggl(\frac{3}{5}\frac{d\ln L}{dt}-\frac{17}{10}\frac{d\ln T}{dt}\biggl) \ ,
\end{align}
where $v_{\rm c}$ is the ejecta velocity at $\Rc$.

When $\Rd>\Rc$, the effective optical depth at $\Rd$ is smaller than unity $\tau_{\rm eff}(\Rd)<1(=\tau_{\rm eff}(\Rc))$ because the optical depth is a decreasing function of radius.
With Eqs. \eqref{eq r_d} and \eqref{eq rho_d} and taking into account the fact that the gas and photon temperatures are the same at $\Rd$, we estimate the effective optical depth by
\begin{align}
\tau_{\rm eff}(\Rd)&\simeq\biggl(\frac{2^4\pi\Ko^2c^5\SB}{3\Kes^4}\biggl)^{1/4}L^{-1/4}T^{-3/4}\vd^{-5/4} \ .
\end{align}
Therefore, the condition $\tau_{\rm eff}(\Rd)<1$ is rewritten as a condition on the velocity \citep[see also][]{Shen+2015}:
\begin{align}
\vd>v_{\rm crit}&\equiv\biggl(\frac{2^4\pi\Ko^2c^5\SB}{3\Kes^4}\biggl)^{1/5}L^{-1/5}T^{-3/5}
	\label{eq v_crit}\\
&\simeq1.9\times10^{4}{\,\rm km\,s^{-1}\,}L_{44}^{-1/5}T_{4}^{-3/5},\nonumber
\end{align}
where we used $\rhod<\rhoc$.
Similarly, we find the condition for $\Rc>\Rd$ to be $\vc<v_{\rm crit}$.
It should be noted that we might not be able to determine $\vd$ and $\vc$ from observations (e.g., using line broadening because different lines are formed at different radii expanding at different velocities).
Thus we determine which radius is larger by using an observed velocity $v$ representative of the outflow velocity.
Unless the velocity profile changes drastically within the ejecta, we determine the situation according to whether $v\gtrless v_{\rm crit}$.

The ejecta mass above the radius $R={\rm max}(\Rd,\,\Rc)$ is obtained by integrating Eq. \eqref{eq mdot d} or \eqref{eq mdot c} over time:
\begin{align}
\Mej(>R)\simeq\int dt \dot{M}(L,T,v) \ .
	\label{eq ejecta mass}
\end{align}
Similarly, the kinetic energy of ejecta is given by
\begin{align}
\Ekin(>R)\simeq\int dt \dot{M}(L,T,v)v^2/2 \ .
	\label{eq kinetic energy}
\end{align}
The upper limit of the integrals should be the time when the radius reaches the center of the ejecta (the beginning of nebular phase).
Although this time is difficult to identify in observations, the results do not depend sensitively on the choice of the time because the mass outflow rate declines after the peak of luminosity (see Figs. \ref{fig SN_sample} and \ref{fig TDE_sample}).

The maximal mass outflow rate is realized when $\Rd=\Rc$ and $\vd=\vc=v_{\rm crit}$ \citep[see also][]{Shen+2015}\footnote{Note that at $v_{\rm crit}$, $f_{\rm d}=f_{\rm c}$.}:
\begin{align}
\dot{M}_{\rm max}&=\biggl(\frac{3^3\pi^2c^5}{2^2\kappa_0\Kes^3\SB^3}\biggl)^{1/5}L^{3/5}T^{-17/10}f_{\rm d}
    \label{eq mdot max}\\
&\simeq6.0\times10^{-1}{\,\Msun\,\rm day^{-1}\,}L_{44}^{3/5}T_{4}^{-17/10}f_{\rm d} \ .\nonumber
\end{align}

\section{Application to supernovae}\label{sec sn}
We apply the  method developed in the previous section to 
a sample of SNe with good data that have been analyzed using detailed numerical simulation \citep{Taddia+2018}.
We compare  our results and verify that our method gives a reasonable estimate of the ejecta mass.

We consider a sample of type Ic SNe. 
For type II SNe, it is well known that the electron scattering opacity decreases drastically due to the hydrogen recombination at the color radius \citep[e.g.,][]{Grassberg+1971,Popov1993,Faran+2019}.
For type IIb and Ib SNe, most helium recombines shortly after the explosion \citep[$\lesssim10\,\rm days$, ][]{Dessart+2011,Piro&Morozova2014} due to its high recombination temperature.
Such an early observation is not always performed and we miss most of the helium ejecta.
 used Thus, in this work, we focus on type Ic SNe for which the ejecta recombination is unimportant and we can ignore the temporal evolution of the electron scattering opacity adopting $\Kes=0.07\,\rm cm^2\,g^{-1}$ \citep[see e.g.,][who estimated the ejecta masses of type Ic SNe using Arnett's rule]{Lyman+2016,Taddia+2018}. The absorption opacity is dominated by bound-bound transitions because the ejecta are mainly composed of metals \citep{Dessart+2015}. Here we use a constant absorption opacity $\Ka=0.01\,\rm cm^2\,g^{-1}$ (see appendix \ref{appendix a}). Note that when discussing SNe, here, we use a different absorption opacity prescription and opacity values than those used in \S \ref{sec method}.

When the ejecta expand homologously after the explosion (as in the case of SNe) we can simplify further the formula for the outflows.
In this case, we can estimate the ejecta velocity at $R$ by $v=R/t$, where $t$ is the time measured since the explosion. We can then rewrite the mass formula using the observed time instead.

For the case of $\Rd>\Rc$, the velocity and outflow rate are given by
\begin{align}
\vd&=\biggl(\frac{3c}{2^4\pi\SB}\biggl)^{1/3}L^{1/3}T^{-4/3}t^{-2/3}
	\label{eq v_d}\\
&\simeq7.5\times10^{4}{\,\rm km\,s^{-1}\,}L_{42}^{1/3}T_{4}^{-4/3}t_{\rm day}^{-2/3} \ ,\nonumber\\
\dot{M}_{\rm d}&=\biggl(\frac{2^23\pi^2c^4}{(\Kes+\Ka)^3\SB}\biggl)^{1/3}L^{1/3}T^{-4/3}t^{1/3}f_{\rm d}
	\label{eq mdot d2}\\
&\simeq1.3\times10^{-1}{\,\Msun\,\rm day^{-1}\,}L_{42}^{1/3}T_{4}^{-4/3}t_{\rm day}^{1/3}f_{\rm d} \ ,\nonumber\\
f_{\rm d}&=\frac{2}{3}\biggl(1-\frac{1}{2}\frac{d\ln L}{d\ln t}+2\frac{d\ln T}{d\ln t}\biggl) \ ,
\end{align}
where $t_{\rm day}=t/\,\rm day$.
For  $\Rc>\Rd$, the velocity and mass outflow rate are given by
\begin{align}
\vc&=\biggl(\frac{3^2(\Kes+\Ka)}{2^8\pi^2\SB^2\Ka}\biggl)^{1/4}L^{1/2}T^{-2}t^{-1}
    \label{eq v_c}\\
&\simeq6.3\times10^{4}{\,\rm km\,s^{-1}\,}L_{42}^{1/2}T_{4}^{-2}t_{\rm day}^{-1}
	\nonumber\\
\dot{M}_{\rm c}&=\frac{3}{2^2\Ka\SB}LT^{-4}t^{-1} f_{\rm c}
    \label{eq mdot c2}\\
&\simeq6.6\times10^{-2}{\,\Msun\,\rm day^{-1}\,}L_{42}T_4^{-4}t_{\rm day}^{-1} f_{\rm c} \ ,
	\nonumber\\
f_{\rm c}&=\biggl(1-\frac{1}{2}\frac{d\ln L}{d\ln t}+2\frac{d\ln T}{d\ln t}\biggl) \ .
\end{align}
Instead of the critical velocity in Eq. \eqref{eq v_crit}, we introduce a critical time before (after) which $\Rd>\Rc$ ($\Rc>\Rd$) by equating $\vd$ with $v_{\rm crit}$ as
\begin{align}
t_{\rm crit}&=\biggl(\frac{3^2(\Kes+\Ka)^3}{2^{8}\pi^2\Ka^3\SB^2c^4}\biggl)^{1/4}L^{1/2}T^{-2}
    \label{eq t_crit}\\
&\simeq6.0\times10^{-1}{\,\rm day\,}L_{42}^{1/2}T_{4}^{-2}\ ,\nonumber
\end{align}
To estimate the velocity accurately, we have to know the moment of the explosion. This is not always well-constrained by the observations, resulting in one of our largest sources of error. 

\begin{figure}
\begin{center}
\includegraphics[width=85mm, angle=0]{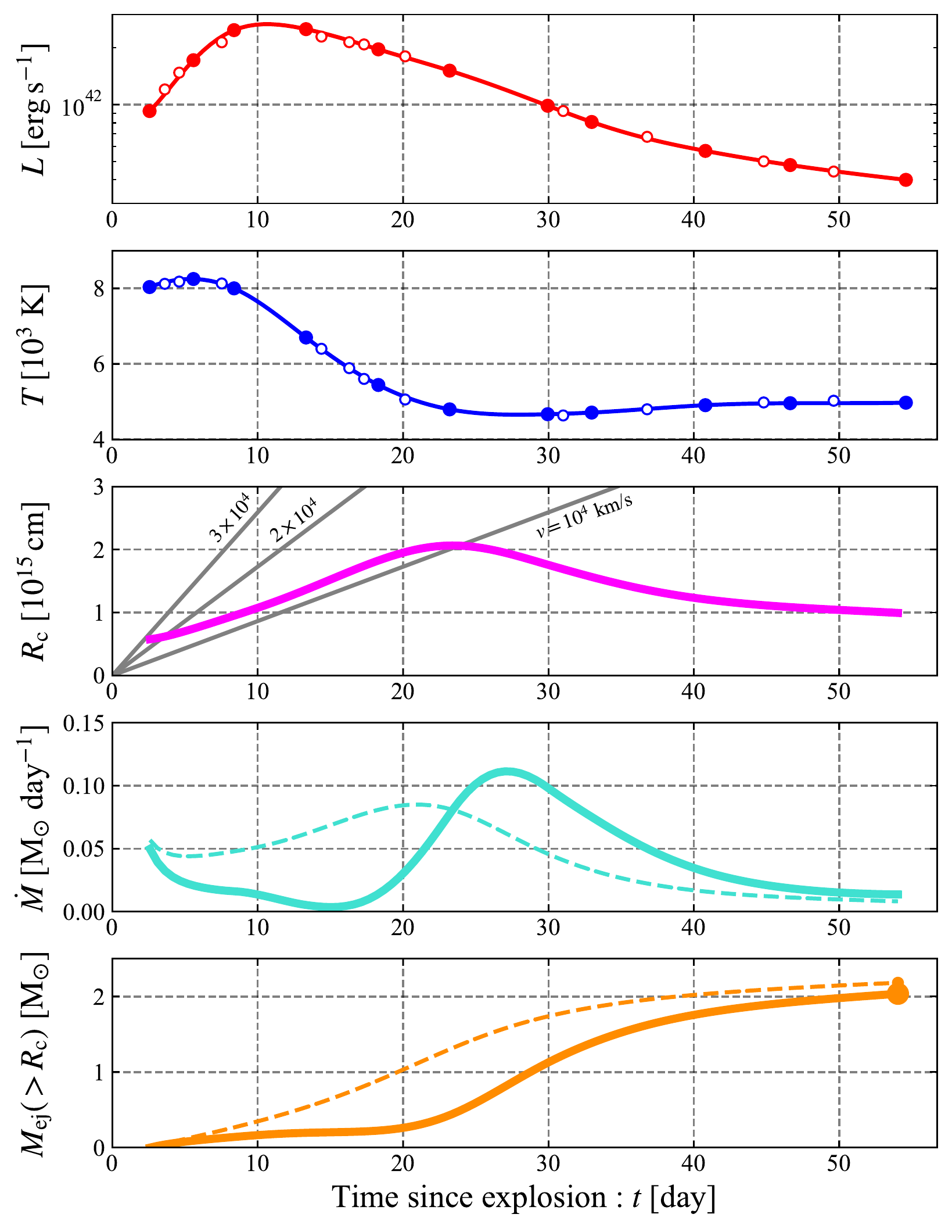}
\caption{The observed luminosity, temperature, estimated color radius $\Rc(>\Rd)$, mass outflow rate, and ejecta mass above the color radius of SN 2004fe.
The observables are taken from figure 12 in \citealt{Taddia+2018}.
In the panel of $\Rc$, the gray lines show the distance of the shells expanding with constant velocities of $v=10^{4}$, $2\times10^{4}$, and $3\times10^4\,\rm km\,s^{-1}$.
For $\dot{M}$ and $\Mej$, we also show the quantities neglecting the term $dR/dt$ ($f_{\rm c}=1$) with dashed curves.
Including the term  $dR/dt$ in Eq. \eqref{eq mdot}, the mass outflow rate is enhanced (suppressed) before (after) the peak of $\Rc$ ($\simeq25\,\rm day$). Note that to obtain smooth fitting curves we used only the solid data points.}
\label{fig SN_sample}
\end{center}
\end{figure}

First we analyze a well-observed SN 2004fe analyzed by \cite{Taddia+2018}.
Fig. \ref{fig SN_sample} depicts the time evolution of the bolometric luminosity, color temperature, calculated color radius, mass outflow rate at $\Rc$, and ejecta mass out of $\Rc$.
For this and the following other SNe, $\Rc$ is always larger than $\Rd$ ($t_{\rm crit}\lesssim\,\rm day$).
The mass outflow rate peaks around $\simeq25\,\rm days$ that reflects the time evolution of $\Rc$.
To see this effect, we also show $\dot{M}$ and $M_{\rm ej}$ neglecting the term $dR/dt$ ($f_{\rm c}=1$) with dashed curves (bottom two panels).
When $\Rc$ expands (shrinks) with time, $\dot{M}$ is suppressed (enhanced) from its value with $f_{\rm c}=1$ (that ignores this effect). This changes $\Mej$ by up to $\sim10\,\%$ (bottom panel).
In Table \ref{table SN_result}, we show the resulting ejecta mass and kinetic energy.
For a comparison, we also show the ratios of our results ($M_{\rm ej},\,E_{\rm kin}$) to the values ($M_{\rm ej,Hy},\,E_{\rm kin,Hy}$) obtained by one-dimensional hydrodynamic simulations with flux-limited diffusion (\citealt{Taddia+2018}, see also \citealt{Bersten+2011,Bersten+2012} for the details of the calculation).
Our results agree with those of the hydrodynamic modeling within a factor of 2.
The errors are calculated by considering the uncertainty of the explosion time (other uncertainties such as the luminosity and temperature are not available in the literature and are neglected here).

\begin{figure*}
\begin{center}
\includegraphics[width=175mm, angle=0]{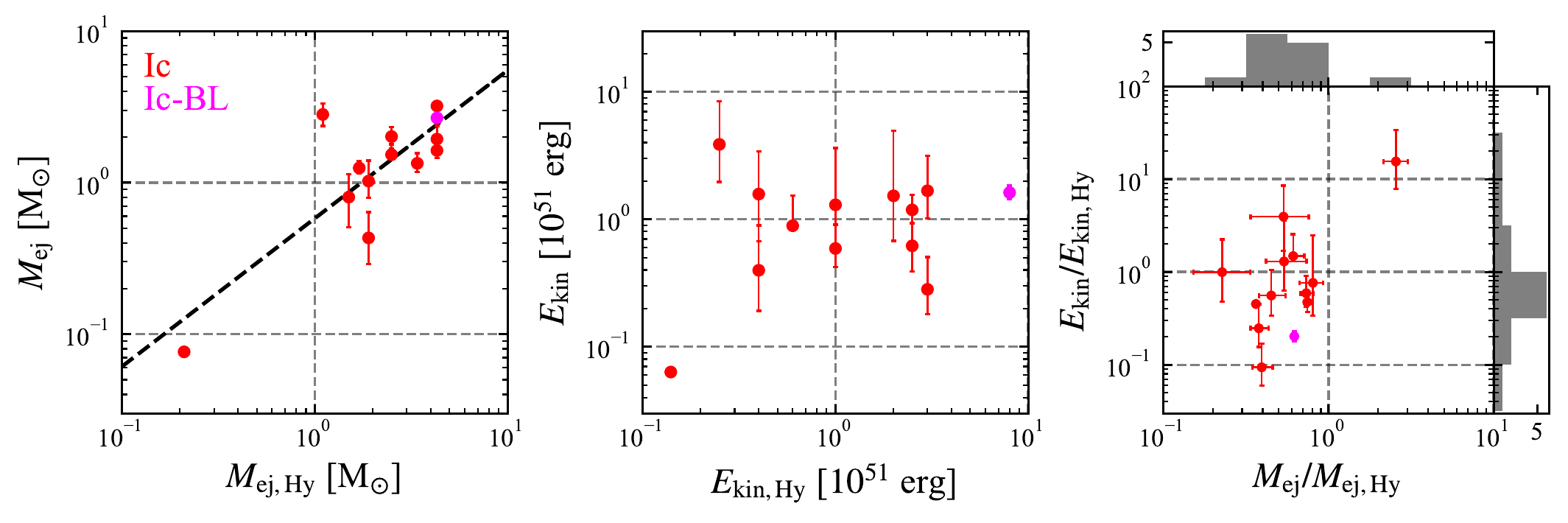}
\caption{({\bf Left}) A comparison of ejecta mass estimated by our method $\Mej$ with those obtained by hydrodynamical modeling, $M_{\rm ej,Hy}$, \citep{Taddia+2018}.
The red and magenta points show the type Ic and broad-line Ic (Ic-BL) SNe, respectively.
The black dashed line shows the best fit to the points: $\log \Mej=0.98\log M_{\rm ej,Hy}-0.24$.
({\bf Middle}) Same as for the left panel but for the kinetic energy.
({\bf Right}) The distribution of the ratios, $\Mej/M_{\rm ej,Hy}$ and $\Ekin/E_{\rm kin,Hy}$.
The mean and standard deviation of the distributions in log space are $-0.24$ and $0.24$ for the mass ratio, and $-0.11$ and $0.56$ for the energy ratio, respectively.}
\label{fig SN_error}
\end{center}
\end{figure*}

We calculate the ejecta mass and kinetic energy of other 11 SNe studied by  \cite{Taddia+2018}\footnote{We exclude SN 2009ca because its luminosity is unusually large among this sample and it might be an outlier.} and one ultra-stripped envelope SN iPTF14gqr \citep{De+2018}.
Fig. \ref{fig SN_error} presents a comparison of our results to those obtained using numerical modeling.
Our estimation of ejecta mass (left panel) is consistent with the hydrodynamic calculation while the kinetic energy distribution (middle panel) shows a discrepancy.
The right panel depicts a histogram of the ratios of quantities obtained by our estimates to those by numerical modeling.
The mean and standard deviation of $\log(M_{\rm ej}/M_{\rm ej,Hy})$ are $-0.24$ and $0.24$, respectively.
Thus our estimation nicely agrees with that by hydrodynamic modeling with an accuracy of a factor 2.
The distribution of $\log(E_{\rm kin}/E_{\rm kin,Hy})$ has a mean and standard deviation of $-0.11$ and $0.56$, respectively.
This poorer agreement is because the kinetic energy is more sensitive to the estimate of the velocity which in turn depends on the explosion time (see error bars in right panel of Fig. \ref{fig SN_error}).

\begin{table}
\begin{center}
\caption{Ejecta mass and kinetic energy of type Ic SNe estimated by our method ($\Mej$, $\Ekin$). We also show their ratios to those obtained by hydrodynamical modeling ($M_{\rm ej,Hy}$, $E_{\rm kin,Hy}$) by \citealt{Taddia+2018}. Only the uncertainty of the explosion time is included to calculate the errors.}
\label{table SN_result}
\begin{tabular}{lrrrr}
\hline
Event &$\Mej$&$\Mej/M_{\rm ej,Hy}$&$\Ekin$&$\Ekin/E_{\rm kin,Hy}$\\
& [$\Msun$]&&[$10^{51}\,\rm erg$]&\\
\hline
SN 2004fe&$2.0_{-0.3}^{+0.3}$&$0.8_{-0.1}^{+0.1}$&$1.5_{-0.9}^{+3.4}$&$0.8_{-0.4}^{+1.7}$\\
SN 2004gt&$1.3_{-0.2}^{+0.2}$&$0.4_{-0.1}^{+0.1}$&$0.3_{-0.1}^{+0.2}$&$0.09_{-0.03}^{+0.1}$\\
SN 2005aw&$1.6_{-0.2}^{+0.2}$&$0.4_{-0.04}^{+0.1}$&$0.6_{-0.2}^{+0.5}$&$0.3_{-0.1}^{+0.2}$\\
SN 2005em&$2.8_{-0.5}^{+0.5}$&$2.6_{-0.4}^{+0.5}$&$3.9_{-1.9}^{+4.6}$&$15.5_{-7.6}^{+18.2}$\\
SN 2006ir&$3.2_{-0.2}^{+0.2}$&$0.7_{-0.03}^{+0.04}$&$1.2_{-0.3}^{+0.4}$&$0.5_{-0.1}^{+0.2}$\\
SN 2007ag&$1.5_{0.0}^{+0.3}$&$0.6_{0.0}^{+0.1}$&$0.9_{0.0}^{+0.6}$&$1.5_{0.0}^{+1.1}$\\
SN 2007hn&$0.8_{-0.3}^{+0.3}$&$0.5_{-0.2}^{+0.2}$&$1.6_{-0.9}^{+1.8}$&$3.9_{-2.3}^{+4.6}$\\
SN 2007rz$^*$&$1.3_{-0.1}^{+0.1}$&$0.7_{-0.1}^{+0.1}$&$0.6_{-0.2}^{+0.3}$&$0.6_{-0.2}^{+0.3}$\\
SN 2008hh&$1.9_{-0.3}^{+0.4}$&$0.5_{-0.1}^{+0.1}$&$1.7_{-0.7}^{+1.5}$&$0.6_{-0.2}^{+0.5}$\\
SN 2009bb&$2.7_{-0.1}^{+0.1}$&$0.6_{-0.02}^{+0.02}$&$1.6_{-0.2}^{+0.2}$&$0.2_{-0.02}^{+0.03}$\\
SN 2009dp&$1.0_{-0.2}^{+0.4}$&$0.5_{-0.1}^{+0.2}$&$1.3_{-0.7}^{+2.3}$&$1.3_{-0.7}^{+2.3}$\\
SN 2009dt&$0.4_{-0.1}^{+0.2}$&$0.2_{-0.1}^{+0.1}$&$0.4_{-0.2}^{+0.5}$&$1.0_{-0.5}^{+1.2}$\\
SN 14gqr$^*$&$0.08$&$0.36$&$0.06$&$0.45$\\
\hline
\multicolumn{5}{l}{$^*$ $M_{\rm ej,Hy}$ and $E_{\rm kin,Hy}$ are not available and we use the values given}\\
\multicolumn{5}{l}{by Arnett's rule \citep{Arnett1980,Arnett1982}.}\\
\end{tabular}
\end{center}
\end{table}

\begin{figure}
\begin{center}
\includegraphics[width=75mm, angle=0]{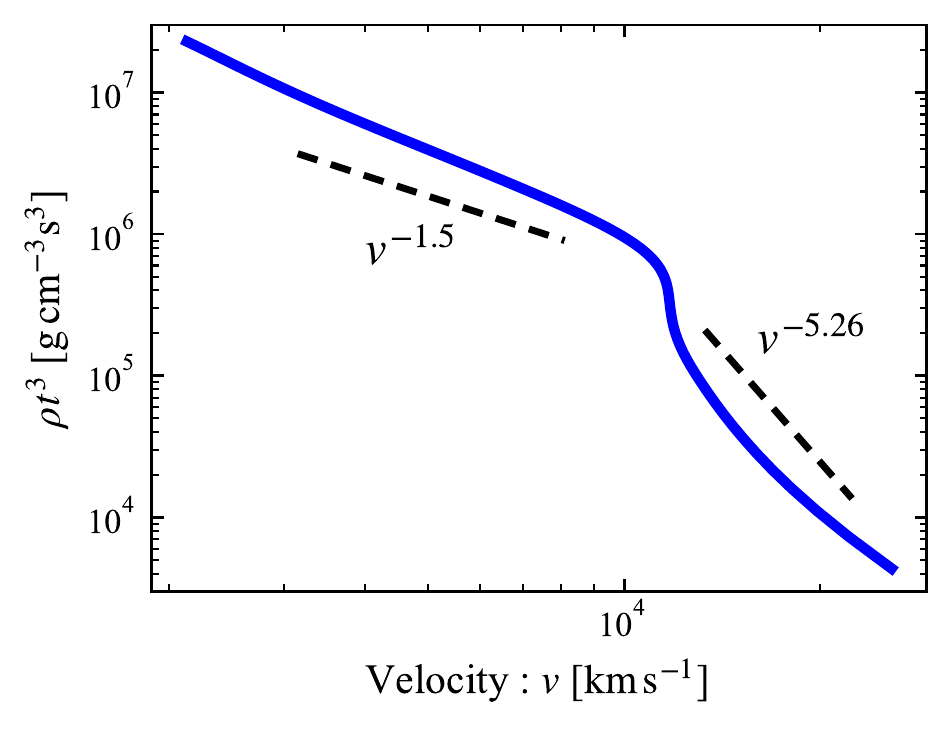}
\caption{Density profile of SN 2004fe in the velocity coordinate reconstructed by correcting the expansion effect ($\rho t^3$).
The profile is described by a broken-power-law function consistent with \citealt{Chevalier&Soker1989,Matzner&McKee1999}.
The inner part has a slope $\rho\propto v^{-1.5}$ and the outer part has a similar power-law index to that expected for Sakurai's solution: $\rho\propto v^{-5.26}$ (\citealt{Sakurai1960}).}
\label{fig SN_density}
\end{center}
\end{figure}

We briefly mention an interesting result of our method concerning  the density profile of SN ejecta.
Since we estimate the density $\rhoc$ for each observation time $t$, we can reconstruct the density profile of the ejecta by correcting the expansion effect $\rho t^{3}$.
Fig. \ref{fig SN_density} depicts a reconstructed density profile of SN 2004fe in the velocity coordinate.
The profile could be described by a broken-power-law and it is consistent with the expected structure \citep{Chevalier&Soker1989,Matzner&McKee1999}.
These works show that the power-law indexes of the inner (slow) and outer (fast) parts are $\sim-1$ and $\sim-10$, respectively.
As shown in Fig. \ref{fig SN_density}, the inner-part's index of $\sim-1.5$ agrees with their results, while the outer part has a shallower slope similar to that given by \cite{Sakurai1960}, $\simeq-5.26$.

\section{Ejecta mass of optical TDEs}\label{sec tde}
We turn now to our original goal estimating the ejecta mass of optical TDEs.
We focus on the ``reprocessing-outflow'' model, where we can assume that a  quasi-spherical wind is launched  and the  thermal photons reprocessed by this wind diffuse out from the outflow \citep{Metzger&Stone2016,Roth+2016,Roth&Kasen2018,Dai+2018,Lu&Bonnerot2020,Piro&Lu2020}.

Consider an optical TDE whose luminosity and temperature are known as a function of time. We integrate  Eqs. \eqref{eq mdot d} and \eqref{eq mdot c} up to the end of the observation time.
We use the electron scattering opacity of $\kappa_{\rm es}=0.35\,\rm cm^2\,g^{-1}$ and the Kramers' absorption opacity with $\kappa_0=4\times10^{25}$ (see Appendix \ref{appendix a}) assuming that the ejecta's chemical composition is solar.
Different from the homologously-expanding SN ejecta, we do not have here a good estimate of the velocity.  As the system is quasi-stationary the outflow may be launched with a constant velocity, we fix the velocity and vary it as a parameter (we will return to this point later).
Similarly, we neglect the term of $dR/dt$ in the mass outflow rate, which becomes $\sim10^{15-16}{\,\rm cm}/100{\,\rm days}\simeq10^{3-4}\,\rm {km\,s^{-1}}$ for a typical event.
This simplification gives us a conservative estimate of ejecta mass (smaller mass) because the radii $\Rd$ and $\Rc$ shrink during observations,  increasing $\dot{M}$.

\begin{figure}
\begin{center}
\includegraphics[width=85mm, angle=0]{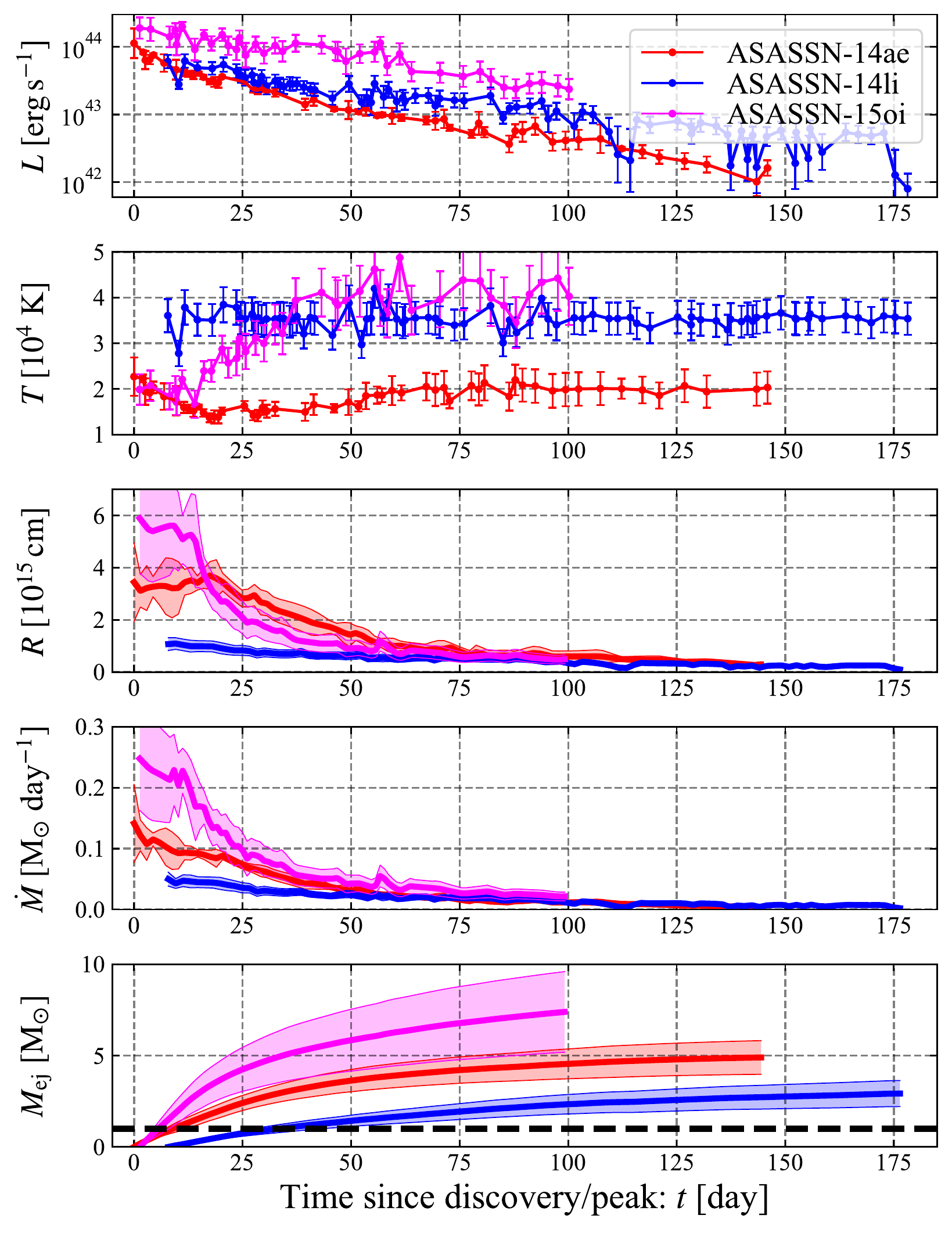}
\caption{Same as Fig. \ref{fig SN_sample} but for optical TDEs ASASSN-14as (red), 14li (blue), and 15oi (magenta) for a fixed velocity of $v=10^{4}\,\rm km\,s^{-1}$. The observables are taken from \citealt{Holoien+2014,Holoien+2016,Holoien+2016b}. 
The shaded regions in the bottom three panels denote the error range calculated when the observational statistical errors in upper two panels is taken into account. The ejecta mass exceeds $\Msun$ (black dashed line), as early as 5-30 days after the discovery. As noticed by \citealt{Piro&Lu2020}, $\Rd$ is larger than the black-body radius $R_{\rm bb}\equiv({L/4\pi\sigma_{\rm SB}T^4})^{1/2}$, that is commonly used to estimate the size of ejecta \citep[see e.g.,][]{Hung+2017}. This is not inconsistent because in the situation we consider  $R_{\rm bb}$ has nothing to do with the emission process \citep[see also][]{Nakar&Sari2010}.}
\label{fig TDE_sample}
\end{center}
\end{figure}

Fig. \ref{fig TDE_sample} depicts the time evolution of the observables ($L$ and $T$), radius, mass outflow rate, and ejecta mass beyond the radius for ASASSN-14ae, 14li, and 15oi (as Fig. \ref{fig SN_sample}) for an assumed fixed velocity of $v=10^4\,\rm km\, s^{-1}$.
In this case, the diffusion radius is always larger than the color radius $\Rd>\Rc$.
The ejecta mass exceeds a solar mass as early as $5-30$ days after the discovery.

\begin{figure*}
\begin{center}
\includegraphics[width=150mm, angle=0]{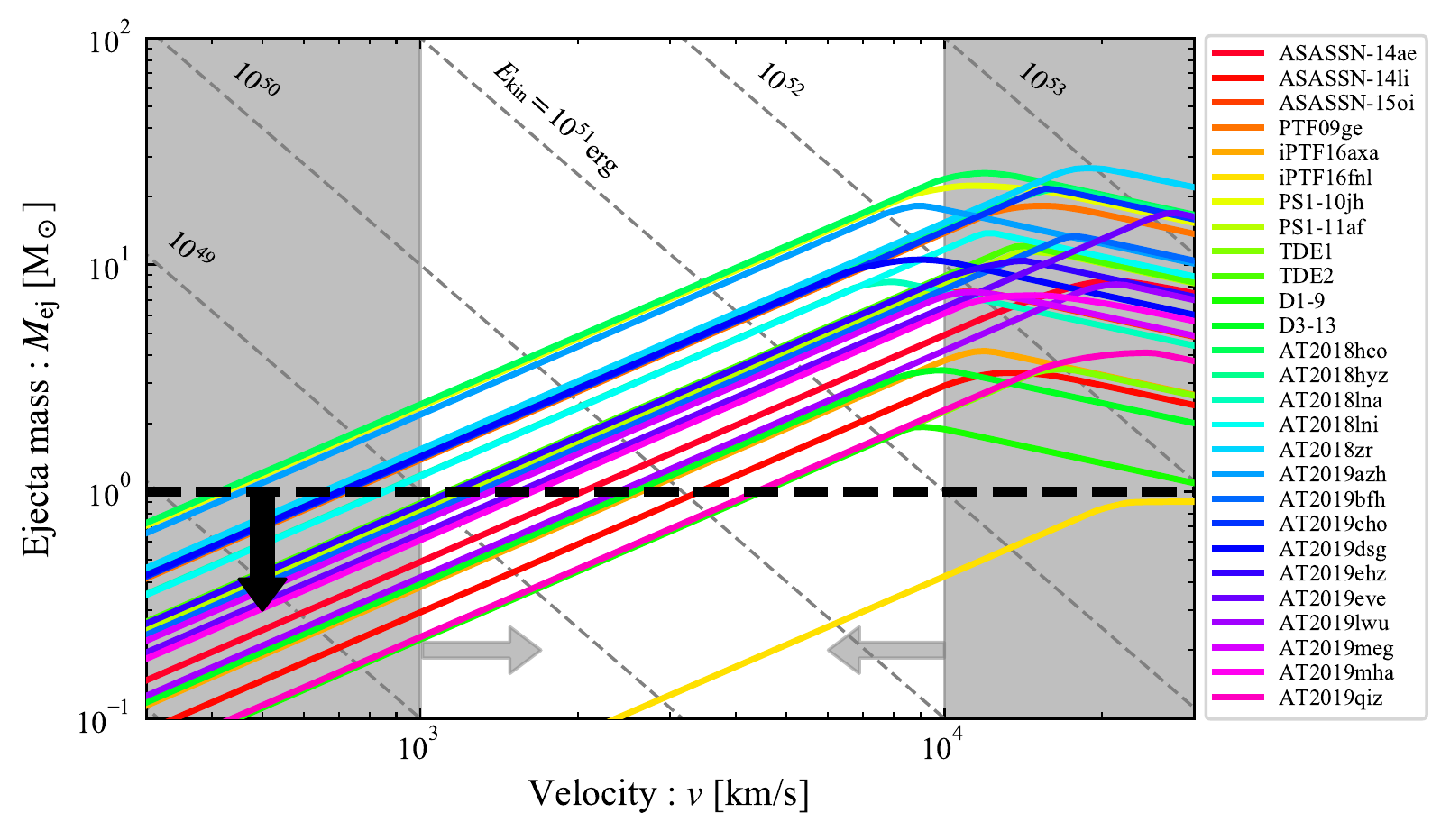}
\caption{Estimated ejecta mass of observed optical TDEs for different ejecta velocities. The gray dashed lines show the kinetic energy at each velocity. As a reference the black dashed line denotes $\Msun$. Typical TDE debris should be bellow this line.
For each event, the ejecta mass has a peak at the critical velocity $v_{\rm crit}\simeq3\times10^3-10^4\,\rm km\,s^{-1}$ (Eq. \ref{eq v_crit}). When $v<v_{\rm crit}$ ($v>v_{\rm crit}$), the color radius is larger (smaller) than the diffusion radius, $\Rd<\Rc$ ($\Rc<\Rd$).
For the ejecta mass to be sufficiently small, the velocity should be smaller than $v\sim10^{2-3}\,\rm km\,s^{-1}$.
The gray shaded regions show velocities excluded by the observations of line widths $v\lesssim10^4\,\rm km\,s^{-1}$ or lower than the escape velocity $v_{\rm esc}\simeq10^3\,\rm km\,s^{-1}$.}
\label{fig TDE_mass}
\end{center}
\end{figure*}

Fig \ref{fig TDE_mass} depicts the ejecta mass of 27 optical TDEs \citep[data are taken from][]{Hung+2017,vanVelzen+2020}\footnote{We exclude AT2018iih which is most likely not a TDE  \citep[see][]{Ryu+2020}. For example, its  luminosity declines much slower than all other TDEs.  The corresponding estimated ejecta mass  $\sim10^2\,\Msun$,  is indeed unreasonably large.}  calculated for different  velocities ranging from $v=3\times10^2$ to $3\times10^4\,\rm km\,s^{-1}$.
The estimated ejecta mass peaks at the critical velocity of $v_{\rm crit}\simeq10^4\,\rm km\,s^{-1}$ for which $\Rd=\Rc$ (Note the dependences of the outflow rate, $\dot{M}\propto v(\propto v^{-1/2})$ for $v<v_{\rm crit}(v>v_{\rm crit})$, see also Eq. \eqref{eq mdot max} and \citealt{Shen+2015}).
Interestingly this critical velocity is comparable to those suggested by line broadening (\citealt{Arcavi+2014} but see also \citealt{Roth+2016,Roth&Kasen2018} for other mechanism giving the observed line broadening) and expected in the ``reprocessing-outflow'' model \citep{Metzger&Stone2016}.
For most  TDEs, the ejecta mass is larger than a solar mass for velocities of $v\simeq3\times10^{3}-10^4\,\rm km\,s^{-1}$ even in our conservative calculation.
The mass is smaller than a solar mass if the ejecta velocity is lower than $v\lesssim10^{2-3}\,\rm km\,s^{-1}$.
Such low velocities lower than the escape velocity are unlikely in the system powered by a super-Eddington disk. Moreover, in this case the ejecta's kinetic energy is as small as $E_{\rm kin}\lesssim10^{49}\,\rm erg$, and this outflow does not solve the ``inverse energy crisis''.

\begin{figure}
\begin{center}
\includegraphics[width=85mm, angle=0]{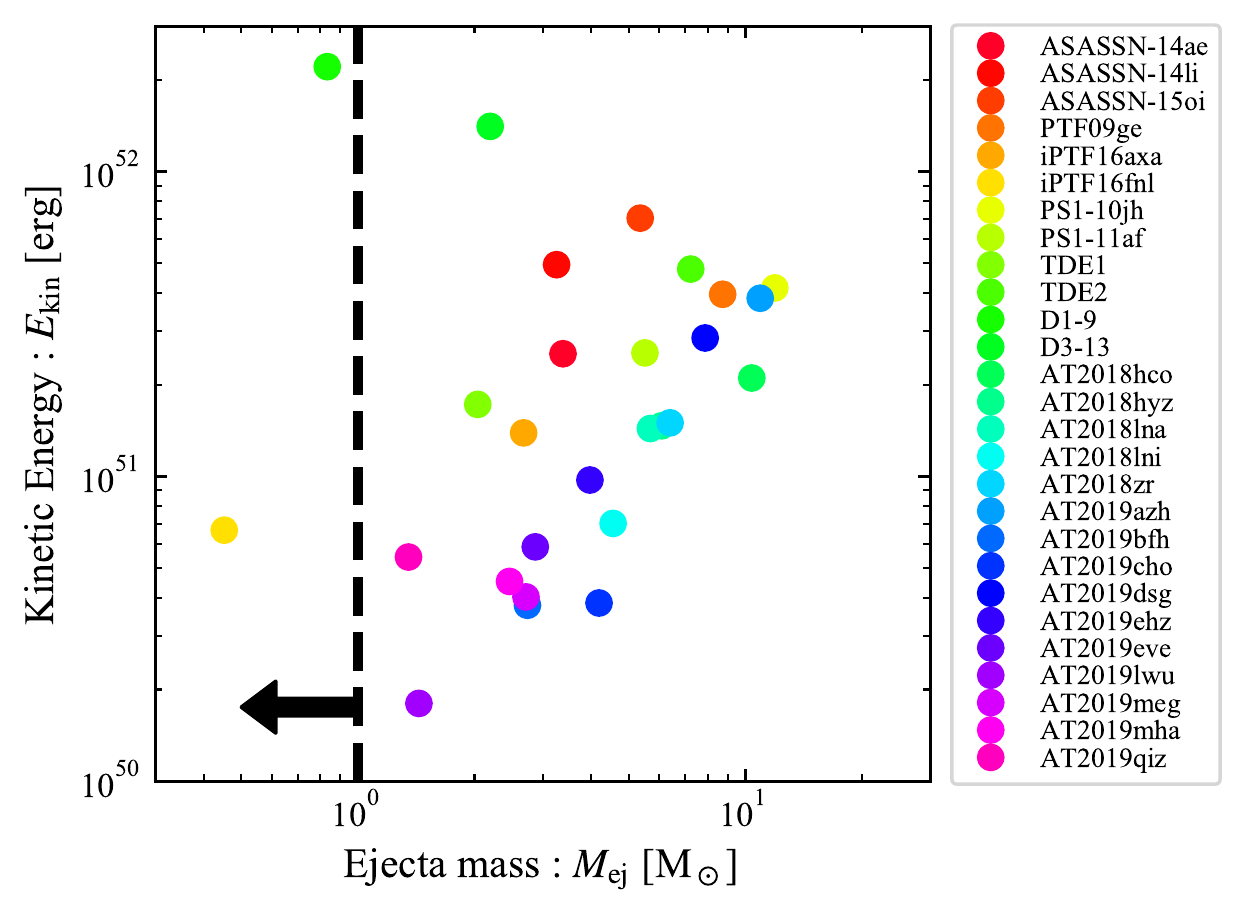}
\caption{The ejecta mass and the kinetic energy calculated by setting the velocity to the escape velocity at $\Rd$ or $\Rc$, $v=v_{\rm esc}(R)$.}
\label{fig TDE_vesc}
\end{center}
\end{figure}

A way to estimate the velocity is adopting the escape velocity at the radius $v_{\rm esc}(R)=\sqrt{GM_{\rm BH}/R}$, which is a lower limit if the sonic radius is inside the radius $<R$ \citep{Shen+2016}.
Here $G$ is the gravitational constant and $M_{\rm BH}$ is the BH mass.
Using Eqs. \eqref{eq r_d} and \eqref{eq r_c}, the escape velocities at radii $\Rd$ and $\Rc$ are; 
\begin{align}
v_{\rm esc}(\Rd)&\simeq\biggl(\frac{2^4\pi \SB G^2M_{\rm BH}^2}{3c}\biggl)^{1/3}L^{-1/3}T^{4/3}
    \label{eq v esc Rd}\\
&\simeq840\,{\rm km\,s^{-1}\,}M_{\rm BH,6.5}^{2/3}L_{44}^{-1/3}T_4^{4/3},\nonumber\\
v_{\rm esc}(\Rc)&\simeq\biggl(\frac{2^{12}\pi^3 \SB^3\kappa_0 G^5M_{\rm BH}^5}{3^3\Kes^2}\biggl)^{1/10}L^{-3/10}T^{17/20}
    \label{eq v esc Rc}\\
&\simeq1900\,{\rm km\,s^{-1}\,}M_{\rm BH,6.5}^{1/2}L_{44}^{-3/10}T_{4}^{17/20},\nonumber
\end{align}
where we use $M_{\rm BH,6.5}=M_{\rm BH}/(10^{6.5}\,\Msun)$.
The resulting typical escape 
velocities are $v_{\rm esc}\sim10^{3}\,\rm km\,s^{-1}$. 
Pluging Eqs. \eqref{eq v esc Rd} and \eqref{eq v esc Rc} into Eqs. \eqref{eq mdot d} and \eqref{eq mdot c} we find  the corresponding mass outflow rates: 
\begin{align}
\dot{M}_{\rm d}&\simeq\biggl(\frac{3^2\pi c^5}{2^2\Kes^3\SB^2GM_{\rm BH}}\biggl)^{1/3}L^{2/3}T^{-8/3}\\
&\simeq2.9\,{\rm \Msun\,day^{-1}\,}M_{\rm BH,6.5}^{-1/3}L_{44}^{2/3}T_{4}^{-8/3},\nonumber\\
\dot{M}_{\rm c}&\simeq\biggl(\frac{3\pi GM_{\rm BH}}{\kappa_0\SB}\biggl)^{1/2}L^{1/2}T^{-1/4}\\
&\simeq1.9\times10^{-1}\,{\rm \Msun\,day^{-1}\,}M_{\rm BH,6.5}^{1/2}L_{44}^{1/2}T_{4}^{-1/4}.\nonumber
\end{align}
Fig. \ref{fig TDE_vesc} depicts the estimated ejecta mass and kinetic energy for $v=v_{\rm esc}(R)$ adopting the BH mass listed in \cite{Hung+2017} or $10^{6.5}\,\Msun$ when it is not available.
The resulting ejecta mass are  larger than a solar mass.

Finally, we comment on the results by \cite{Piro&Lu2020}.
These authors updated the formalism of \cite{Metzger&Stone2016}.
Assuming density and temperature profiles, they reproduced a TDE light curve with ejecta mass of $\Mej=0.5\,\Msun$.
However, they obtained a luminosity of $\sim10^{43}\,\rm erg\,s^{-1}$ and temperature $\sim10^{5}\,\rm K$ (see their figure 5), which are inconsistent with those observed for typical TDEs: $\sim10^{44}\,\rm erg\,s^{-1}$ and $\simeq3\times10^4\,\rm K$.
Using such a low luminosity and high temperature we also obtain an ejecta mass that is less than a solar mass.

\section{Summary}\label{sec summary}
We present here a new method, based on  earlier ideas of \cite{Shen+2015} and \cite{Piro&Lu2020}, to estimate the ejecta mass of optical transients that involve a quasi-spherically expanding outflow and have a thermal spectrum. 
To test the method, we calculated the ejecta mass of type Ic SNe and confirmed that it gives a reasonable estimate with an accuracy of a factor 2.
Interestingly, for a well-observed SN we can also explore  the velocity and density structure of the ejecta and  we find a density profile consistent with the expected structure \citep{Sakurai1960,Chevalier&Soker1989,Matzner&McKee1999}.

Assuming the ``reprocessing-outflow''  model in which TDEs are powered by a compact source (accretion disk) whose soft X-ray emission is reprocessed at a larger radius to optical/UV emission \citep{Metzger&Stone2016,Roth+2016,Roth&Kasen2018,Dai+2018,Lu&Bonnerot2020,Piro&Lu2020} we apply this  method to optical TDEs.
Unlike SNe the outflow is not expanding homologously and it is in a quasi-steady state. As a results the velocities are unknown. 
While observations show line broadening corresponding to $v\sim10^4\,\rm km\,s^{-1}$ \citep[e.g.,][]{Arcavi+2014}, it is not clear that this reflects the outflow velocity \citep{Roth+2016,Roth&Kasen2018}.
Hence, we assume different ejecta velocities and explore the dependence of the  results on the adopted velocity. As we expect a quasi-steady state outflow, the assumption of a constant velocity is reasonable. Alternatively following hydrodynamic modeling by \cite{Shen+2016} we use the escape velocity.

We find reasonable ejecta masses (less than $\sim$solar) only for ejecta velocities less than about thousand $\rm km\,s^{-1}$.
The corresponding ejecta kinetic energy is smaller than $10^{49}\,\rm erg$, which is well below the expected value ($\sim10^{52-53}\,\rm erg$) if the ejecta are launched from the compact source and does not resolve the inverse energy crisis. 
For larger velocities comparable to those inferred from the line width or implied by the escape velocity,  $v\simeq3\times10^3-10^4\,\rm km\,s^{-1}$, the ejected mass is significantly larger than  solar. 
These results cast some doubt on the overall picture of a wide range of ``reprocessing-outflow'' model.
We find that the resulting wind would be either too massive or if it is not massive it will not carry sufficient energy to provide the missing ``energy sink''. 

Some reprocessing models \citep{Loeb&Ulmer1997,Guillochon+2014,Coughlin&Begelman2014,Roth+2016} assume a quasi-static layer of poorly circularized debris reprocesses the ionizing continuum from the (efficiently circularized) inner accretion disk. As the assumed velocities are small in this case the ejecta mass estimates do not pose a problem. However, the ``inverse energy crisis'' remains here as it is not clear how does the excess energy disappear.

There are several caveats in our results.
The first is the assumption of spherical geometry.
Clearly the configuration is not spherical. However, while such deviations are natural in a TDE, we expect that if a deviation is on a large angular scale it could be taken into account by considering the fraction of the solid angle subtended by the outflow. This will not change quantitatively our results. 
We expect that only large deviations such as a formation of a jet would  change our conclusions significantly. \cite{VanVelzen+2013} use late-time radio observations to put an upper limits on Sw J1644 like events (jetted TDEs). They find that  at high probability all seven events that they analyzed did not have $>10^{52}$ erg jets. A more detailed analysis would likely yield much stronger limits.

A second caveat is the assumption that the emission is thermalized and is well described by a single temperature black-body. While this is natural if the emission is indeed reprocessed by optically thick matter, lacking a detailed measurement of the spectrum it is still unclear. Moreover, 
the temperature is usually determined  only  by using  multi-band photometry. If for the given luminosity the true temperature is larger by a factor of 3 the ejecta mass would decrease by a factor of $\sim 3- 10 $ (see Eqs. \ref{eq mdot d} and \ref{eq mdot c}), leading possibly to acceptable solution. However, if the luminosity is also varied to reflect the higher temperature the mass estimate remains valid.  

To conclude, we presented here a simple generic  method to estimate the outflow from quasi-spherical optical transients. The method is valid for events that are optically thick and when the luminosity and temperature have been measured as a function of time. To test the method we have  applied it to SNe finding an encouraging agreement with more detailed calculations. When applying it to optical TDEs, we find that if indeed the observed emission is thermal and it arises from a compact accretion disk then the resulting mass of the outflow is too large. This poses a problem for the ``reprocessing-outflow'' model. 
This problem does not arise in models in which an outflow is not expected such as  the alternative ``outer shocks'' and in variants of the reprocessing model considering a static envelop. 
With numerous on-going and forthcoming observational campaigns aiming at exploring the transient universe it will be interesting to analyze in future work other optical transients using this method.

\section*{acknowledgments}
We thanks Iair Arcavi, Chi-Ho Chan, Tamar Faran, Julian Krolik, Keiichi Maeda,  Kartick Sarkar, Nicholas Stone, and Masaomi Tanaka for fruitful discussions and helpful comments.
We also appreiate an anonymous referee for helpful suggestions.
This work is supported in part by JSPS Postdoctral Fellowship, Kakenhi No. 19J00214 (T.M.) and by ERC advanced grant ``TReX'' (T.P.).

\section*{data availability}
The data underlying this article will be shared on reasonable request to the corresponding author.

\appendix 
\section{Absorption opacity}\label{appendix a}
\begin{figure}
\begin{center}
\includegraphics[width=85mm, angle=0]{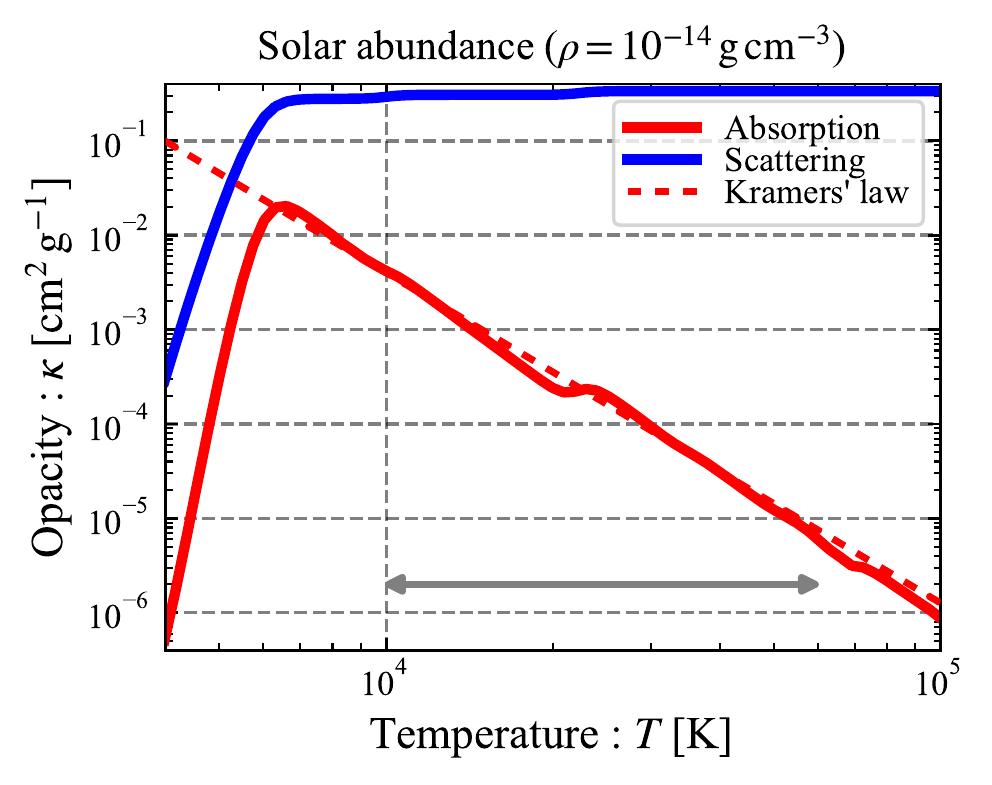}
\caption{Absorption and scattering opacities for solar abundance ejecta (relevant to TDEs) calculated using {\tt CLOUDY}. The absorption opacity is calculated by averaging frequency-dependent continuum (free-free and bound-free) opacity with Planck function (Planck-mean). The dashed line shows the Kramers' opacity adopted in our calculation, which well approximates the absorption opacity in the TDE temperature range (gray arrow).}
\label{fig kappa_solar}
\end{center}
\end{figure}

The Planck-mean opacity is defined by
\begin{align}
\kappa_{\rm a}=\frac{\int_0^\infty \kappa_{\rm a,\nu}B_{\nu}(T)d\nu}{\int_0^\infty B_\nu(T)d\nu},
\end{align}
where $\kappa_{\rm a,\nu}$ and $B_\nu(T)$ are the frequency-dependent absorption opacity and the Planck function, respectively.
We calculate $\kappa_{\rm a,\nu}$ using the open-source code \texttt{CLOUDY}.\footnote{Version 17.01 of the code last described \citep{Ferland+2017}.}
For TDE ejecta, we include only continuum opacity (free-free and bound-free absorption) because for TDE temperature $\gtrsim10^4\,\rm K$ the bound-bound transitions are typically unimportant.  
Figs. \ref{fig kappa_solar} depicts the resulting Planck-mean opacity for TDEs (solar abundance). The absorption opacity is well approximated by the Kramers' opacity formula $\kappa_{\rm a}=\kappa_0\rho T^{-7/2}$ with the normalization of $\kappa_0=4\times10^{25}$. Note that this value is about 10 times larger than the one adopted to approximate the Rosseland-mean opacity.

For type Ic SN ejecta, bound-bound transitions may dominate the absorption opacity whose dependence on density and temperature is less clear.
In analytical light-curve modeling of type Ia SNe, a constant opacity is usually adopted \citep[e.g.,][]{Piro&Nakar2014}.
\cite{Kasen+2013} present a sample of opacity calculation for silicon, which shows that the opacity dose not depend on temperature sensitively  (see their figure 6). Following the calculation by \cite{Dessart+2015} (see their figure 24) we adopt here a constant Planck-mean opacity with $\Ka=0.01\,\rm cm^2\,g^{-1}$.

\bibliographystyle{mnras}
\bibliography{reference_matsumoto}

\bsp	
\label{lastpage}
\end{document}